\documentclass[preprint2]{emulateapj}
\usepackage{graphicx}
\usepackage{natbib}
\usepackage[dvipsnames]{color}
\usepackage{amsmath}

\def \tr{\textcolor{red}}

\def\simless{\mathbin{\lower 3pt\hbox
   {$\rlap{\raise 5pt\hbox{$\char'074$}}\mathchar"7218$}}}
\def\simgreat{\mathbin{\lower 3pt\hbox
   {$\rlap{\raise 5pt\hbox{$\char'076$}}\mathchar"7218$}}}

\def\simless{\mathbin{\lower 3pt\hbox
   {$\rlap{\raise 5pt\hbox{$\char'074$}}\mathchar"7218$}}}
\def\simgreat{\mathbin{\lower 3pt\hbox
   {$\rlap{\raise 5pt\hbox{$\char'076$}}\mathchar"7218$}}}

\begin{document}

\title{Does the chemothermal instability have any role in the 
fragmentation of primordial gas}
\author
{Jayanta Dutta}

\affil{
Department of Physics, Indian Institute of Science, Bangalore-560012, 
India
\break email: jd.astrop@gmail.com \\
}

\begin{abstract}
The collapse of the primordial gas in the density regime 
$\sim 10^{8}\hbox{--}10^{10}$ cm$^{-3}$ is controlled by the three-body 
$\rm H_2$ formation process, in which the gas can cool faster than  
free-fall time $\hbox{--}$ a condition proposed as the chemothermal 
instability. We investigate how the heating and cooling rates are 
affected during the rapid transformation of atomic to molecular 
hydrogen. With a detailed study of the heating and cooling balance 
in a 3D simulation of Pop~III collapse, we follow the chemical and 
thermal evolution of the primordial gas in two dark matter minihaloes. 
The inclusion of sink particles in modified Gadget-2 smoothed particle 
hydrodynamics code allows us to investigate the long term evolution of 
the disk that fragments into several clumps. We find that the sum of 
all the cooling rates is less than the total heating rate after 
including the contribution from the compressional heating ($pdV$). 
The increasing cooling rate during the rapid increase of the molecular 
fraction is offset by the unavoidable heating due to gas contraction. 
We conclude that fragmentation occurs because $\rm H_2$ cooling, 
the heating due to $\rm H_2$ formation and compressional heating 
together set a density and temperature structure in the disk that 
favors fragmentation, not the chemothermal instability.
\end{abstract}


\maketitle

\section{Introduction}
\label{sec:introduction}
Understanding the first source of light in the Universe, the so-called 
primordial stars or Population III (Pop~III) stars, is crucial in 
determining how the Universe evolved into what we observe today 
(\citealt{by11,b13,g13}). With the help of state-of-the-art simulations 
along with the well established cosmological parameter \citep{planck2014}, 
recent studies have provided a standard model of the Pop~III star 
formation (e.g., \citealt{gbcgskys12,jdk13,sgkbl13,hirano14,hirano15,hcgks15}).
In this picture, the baryons (mainly atomic hydrogen) in low mass 
($\sim 10^5\hbox{--}10^6$ $M_\odot$) dark matter halo with virial 
temperature $\sim 1000$ K gravitationally collapsed at redshift 
$z \geq$ 20 via $\rm H_2$ molecular cooling to form the very first 
stars in the Universe (\citealt{htl96,tegmark97,bl01,bl04,cf05}).

Initially the gas is cooled via $\rm H_2$ rotational and vibrational 
line emission \citep{suny98,on98}. However, the quick conversion 
of atomic hydrogen into molecules via the three-body reactions 
\citep{pss83}: 
\begin{equation}
\rm H + H + H \rightarrow H_2 + H
\end{equation}
\begin{equation}
\rm H + H + H_2 \rightarrow H_2 + H_2
\end{equation}
can cool the gas rapidly at high densities ($\sim 10^8$ cm$^{-3}$). 
This results in the gas to be chemothermally unstable 
\citep{yoha06,tao09}. The physical processes are extremely complex 
during the three-body reaction. The gas undergoes both heating 
and cooling simultaneously. Heating is due to the release of 
4.4 eV energy associated with every $\rm H_2$ formation and gas 
contraction, and cooling is due to the emission, dissociation 
of $\rm H_2$ and collisions of $\rm H_2$. However, there is 
little understanding of the interplay between the heating and cooling 
rates so far.

In addition, the physics at high densities is relevant to 
the inception of the thermal instability \cite[first predicted 
by e.g.,][]{sy77,silk1983} that can generate fluctuations in the gas 
density in a comparatively shorter interval than the collapse time. 
The numerical simulations by \citet{abn02} found several 
chemothermally unstable regions. They have, however, pointed out that 
the instabilities do not lead to fragmentation as the turbulence 
efficiently mixes the gas, thereby erasing the fluctuations 
before they can grow significantly. In a different approach, 
\citet{oy03} have calculated the stability of the gas cloud for 
isobaric perturbations by introducing the `growth parameter ($Q$)', 
which must be significantly larger than unity for a clump of gas 
to break into multiple objects. A detailed analytical calculation of 
the instability criterion has been investigated by \citet{ra04}. 
The high resolution 3D $\Lambda$CDM cosmological simulation by 
\citet{yoha06} has shown that the growth parameter indeed never 
becomes much larger than unity. It is infact less than 1.5, 
which implies that the chemothermal instability cannot lead 
to any fragmentation. A similar conclusion was drawn by \citet{tao09}. 
In contrast, more recently \citet{gsb13} have used {\em Arepo} 
simulations to follow the evolution of the gas in minihaloes that 
fragments. They have suggested that this is due to the 
chemothermal instability because the ratio of the cooling 
timescale to the free-fall timescale drops below unity 
in the density space where the three-body reaction dominates.

Our approach to study the chemothermal instability as the sole 
reason for fragmentation scenarios is significantly different from
those in the previous works. For example, for the first time to the 
best of our knowledge, we investigate the heating and cooling balance 
in a full 3D simulation of Pop~III collapse to understand better the 
physical processes during the rapid conversion of atomic to molecular 
hydrogen. The closest study so far would be the analysis by \citet{om05}. 
However, they examined idealised 1D collapse models without
exploring the uncertainties in the three-body rates. In this 
paper, we investigate the chemothermal instability by comparing 
different heating and cooling rates and discuss their roles in the 
fragmentation during the long-term evolution of the disk.

\section{Numerical Methodology}
\label{sec:simulation}

The simulation setup and initial conditions are similar to our previous
study \citep{dnck15} that used minihaloes from the cosmological simulations 
of \cite{gswgcskb11}. Both the haloes start with a maximum central cloud 
number density $n \sim$ 10$^{6}$ cm$^{-3}$, the density before the onset 
of the three-body reaction. The details of the halo properties are given 
in Table~1. We use the halo configurations as the initial condition for
our standard SPH code Gadget-2 \citep{springel05} that has been modified 
with the inclusion of sink particles \citep{bbp95,jap05} and time-dependent 
chemical network for primordial gas \citep{cgkb11a}. The snapshots 
from the hydrodynamic moving mesh code {\em Arepo} \citep{springel10} 
are converted into the Gadget-2 implementation \citep{sgcgk11}. This is 
possible because the mesh-generating points of {\em Arepo} can be 
interpreted as the Lagrangian fluid particles .
The mass resolution in our Gadget-2 simulation is 
$\approx$ 10$^{-2}$ $M_\odot$ for 100 SPH particles \citep{bb97}.

Once the gas density reaches $\sim 10^{10}$ cm$^{-3}$, the cloud 
becomes opaque and the strongest of $\rm H_2$ lines becomes optically 
thick. The $\rm H_2$ cooling rate in this regime is calculated 
using the Sobolev approximation (as described in \citealt{yoha06}). At 
densities $\geq 10^{14}$ cm$^{-3}$, the gas goes through a phase of 
cooling instability due to rapid increase in the cooling rate by 
$\rm H_2$ collisional induced emission (CIE). We follow \citet{ra04} 
to use the rate in our cooling function. Above the central density 
$\sim 10^{16}$ cm$^{-3}$, the gas becomes completely optically thick 
to the continuum radiation \citep{yoh08}. At this point, the 
remaining $\rm H_2$ dissociates by collisions with the atomic $\rm H$ 
and other $\rm H_2$ molecules, and consequently cools the gas 
resulting in further collapse.

We ensure that our investigation is not biased on the chemical 
uncertainties and halo configuration. Hence, we simulate the 
gas evolution in two different minihaloes with the extreme three-body 
rate coefficients: \citet{abn02} (hereafter ABN02) provide the slowest 
rate, \citet{fh07} (hereafter FH07) provide the fastest rate. 
Using the highly simplified one-zone models, the works by \citet{ga08} 
and \citet{gs09} examined minutely the effects of the uncertainty in 
the three-body $\rm H_2$ formation rates and the cooling rate on the 
thermal evolution of the collapsing gas in simple Bonnor-Ebert spheres 
\citep{ebert1955,bonnor1956}. They have found that the large 
uncertainty between the ABN02 and FH07, which differ by a large amount, 
lead to an uncertainty of approximately 50\% in the temperature 
evolution of the gas in the density range $10^8 < n < 10^{13}$ cm$^{-3}$.

The primordial protostar can be modeled as a sink particle \citep{kmk04},
which basically replaces the high-density region as a protostar that can 
accrete infalling mass. In our simulations, the protostar is formed 
once the number density of the gas reaches 5 $\times 10^{13}$ cm$^{-3}$ 
and temperature is $\sim$ 1000 K. The Jeans mass at this critical point 
is $M_{\rm J}$ (1000\,K, $10^{13}\,$ cm$^{-3}$) $\sim$ 0.06 $M_\odot$ 
and Jeans radius is 6 AU, which is the accretion radius, 
$r_{\rm acc}$, of the sink particle. The spurious formation of new sink 
particles is avoided by preventing the sink particles from forming within 
$2\, r_{\rm acc}$ of one another. Depending on the halo configurations 
and chemical uncertainties, our simulations run $\approx 3500\hbox{--}6500$ 
yr after the formation of the first protostar.
 

\begin{table}
\label{tab:3bh2rates}
\centering
  \begin{tabular}{|c|c|c|}
    \hline               
Halo                &               halo1              &              halo2          \\
properties          &                                &                           \\ 
\hline
\hline                               
 n (cm$^{-3}$)      &  10$^{6}$ (max) 71 (min)        &  10$^{6}$  (max)  85 (min) \\
\hline 
 T (K)              &  469 (max)      59 (min)        &   436 (max)       54 (min) \\ 
\hline
 mass ($M_{\odot}$) &  1030                          &  1093                     \\ 
\hline
 n-SPH              &  690855                        & 628773                    \\ 
\hline
 resolution ($M_{\odot}$)&  1.3 $\times 10^{-2}$     &  1.4 $\times 10^{-2}$     \\
 for 100 n-SPH           &                           &                           \\ 
\hline
  \end{tabular}
\caption{Summary of minihalos from the cosmological simulations. n-SPH
stands for the number of SPH particles in the simulation.}
\end{table}

%
%
%
%
%
%
%
%
%
%
%

\section{Results}
\subsection{Chemothermal instability}
\label{sec:instability}
\begin{figure*}
\centerline{
\includegraphics[width=6.0in]{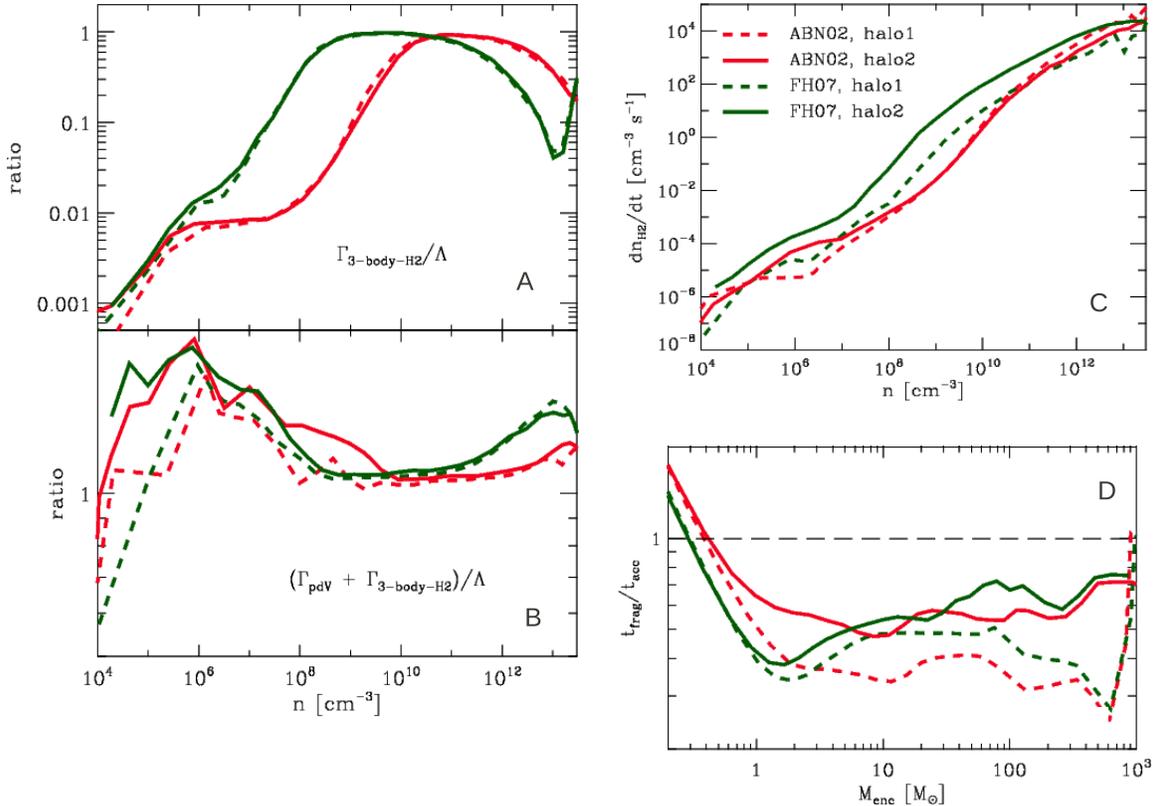}
}
\caption{\label{fig:gas_props} 
Radially binned, mass-weighted averages of the physical quantities 
for different proposed three-body $\rm H_2$ formation rates (red: 
ABN02, green: FH07) of two cosmological halos (dotted: halo1, solid: 
halo2) are compared just before the first protostar is formed. (A) 
The ratio of the three-body formation heating to the total cooing 
rate, (B) ratio of the total heating to the total cooling rate and 
(C) the rate of $\rm H_2$ formation $(dn_{\rm H_2}/dt)$ from the 
simulations in the units of cm$^{-3}$ s$^{-1}$ are plotted as a 
function of density. (D) The fragmentation time over the accretion 
time is plotted as a function of enclosed gas mass.}
\end{figure*}

As the gas undergoes gravitational collapse, the crucial aspect 
to be investigated is whether the gas will be able to cool 
significantly or not. The rapid formation of hydrogen molecules 
via three-body reactions allows the gas to cool in less than 
a free-fall time, heralding chemothermal instability. This effect 
for both the ABN02 and FH07 rates has been shown by \citet{dnck15}, 
where we have found that the ratio of the cooling time to the 
free-fall time ($t_{\rm cool}/t_{\rm ff}$) dips below unity at 
a certain density. This is exactly what \citet{gsb13} 
have found in their simulations. In our present 
calculation, we have considered all the cooling mechanisms, 
including the contributions from chemical heating, to calculate 
the cooling time: 
$t_{\rm cool} = \epsilon / \Lambda_{\rm net}$, where $\epsilon$ 
is the energy per unit volume of the gas and $\Lambda_{\rm net}$ 
is the net cooling rate in units of erg s$^{-1}$ cm$^{-3}$. This 
implies that the gas can cool as it collapses, and is hence 
undergoing chemothermal instability. 
It is to be noted that the drop in the ratio 
$t_{\rm cool} / t_{\rm ff}$ coincides with the onset (in density 
space) of the three-body $\rm H_2$ formation \citep{gsb13,dnck15}. 

However, as we have discussed in Section~\ref{sec:introduction}, 
there is still some discrepancy of whether the fragmentation is 
stimulated by the chemothermal instability or not. Fragmentation 
might also depend on other details, such as turbulence, rotation, 
etc (see for examples, \citealt{tao09,sgb10,cgsgkb11b,latif15,dutta15a}). 
The variation of the ratio of the cooling time to the free-fall 
time in the density space, where the three-body reactions take place, 
is too simple to make a definite conclusion about the 
fragmentation. A straightforward analysis involves comparing all 
possible heating and cooling processes that are responsible for 
setting the thermal balance in the gas.

\subsection{Heating and Cooling rates}
\label{sec:rates}

In this section, we investigate the relevant cooling and heating 
mechanisms associated with the emission, chemical reaction 
and gas contraction during collapse. The synchronous effects of 
the cooling and heating rates make the chemical and 
thermodynamical evolution complicated

Assuming the evolution of the gas density ($\rho$) with the 
free-fall time ($t_{\rm ff}$), i.e., $d\rho/dt = \rho/t_{\rm ff}$, 
previous studies (e.g., \citealt{om2000,bcl02}) show that the 
thermal evolution is followed by solving the energy equation:
\begin{equation}
\frac{d\epsilon}{dt} = \frac{p}{\rho}\frac{d\rho}{dt} - \Lambda + \Gamma ,
\end{equation}
where 
$\Lambda$ and $\Gamma$
are the cooling and heating rate respectively in units of 
erg s$^{-1}$ cm$^{-3}$. This can be seen in Figure~\ref{fig:gas_props} 
in which we compare the various heating and cooling rates in the 
simulation as a function of density. All the quantities are 
mass-weighted averages of the individual SPH particles that are 
binned logarithmically in radius. These logarithmic-binned 
spherical shells are centered at the origin at $r=0$. We have 
rigorously verified that the mass-weighted averaging does not 
throw away any physical information. The calculations are 
done once the central region has collapsed to a density 
$\sim$ 5 $\times 10^{13}$ cm$^{-3}$, i.e., just before the formation 
of the first sink.  

We take the total cooling rate as the sum of the $\rm H_2$ 
line-cooling rate, collision induced emission cooling rate (CIE) 
and dissociation rate (i.e. 
$\Lambda = \Lambda_{\rm line} + \Lambda_{\rm CIE} + \Lambda_{\rm Diss}$) 
and total heating rate as the combination of heating from the 
compression ($pdV$) and the three-body formation of $\rm H_2$ 
(i.e. $\Gamma = \Gamma_{\rm pdV} + \Gamma_{\rm 3-body-H2})$.

Figure~\ref{fig:gas_props}A shows that the total cooling can be 
as high as the heating associated with the three-body formation 
of $\rm H_2$, and thus the ratio of $\Gamma_{\rm 3-body-H2}$ 
to $\Lambda$ is of order unity in the density space where 
the three-body reaction dominates. The three-body heating 
rates for the FH07 rise before the ABN02 do because the higher 
$\rm H_2$ formation rates of FH07 cause the gas to become fully 
molecular at lower densities. As a result, the gas in the FH07 
case becomes optically thick at earlier times, preventing the 
thermal energy from being transported efficiently out of the disk.
We have verified that other cooling processes, such as the heat 
loss due to the collisional dissociation or the collision 
induced emission are not effective until much higher densities.

Most importantly, when including the contribution from compressional 
heating, we notice that the total heating rate is always greater 
than the total cooling rate of the gas (Figure~\ref{fig:gas_props}B). 
For the gas to be chemothermally unstable to fragmentation, the 
heating and cooling rate must satisfy the condition: 
\begin{equation}
\Lambda_{\rm line} + \Lambda_{\rm CIE} + \Lambda_{\rm Diss} \geq \Gamma_{\rm pdV} + \Gamma_{\rm 3-body-H2} \, , 
\end{equation} 
that could occur due to a sharp increase in the fractional abundances. 
However, in our case, the total heating rate dominates even in the 
high density regime. This leads to an ever increasing temperature 
with density. 

From this analysis, we tentatively conclude that the gas can 
experience a dip in the ratio of the cooling time to the free-fall 
time in the density space, and hence becomes chemothermally unstable. 
However, the rapid cooling due to the three-body reaction is 
countervailed by the compressional heating. Thus it never 
leads to significant drop in the temperature. This is consistent 
with the previous studies (e.g., \citealt{abn02,bcl02,ra04,yoha06}).

At this point, we would like to point out the differences with 
the  work by \citet{gsb13}. The use of the ratio 
$t_{\rm cool}/t_{\rm ff}$ as a criterion for fragmentation and 
collapse by \citet{gsb13} neglects the heating due to the formation 
of $\rm H_2$ and $pdV$ contraction. This heating generally exceeds 
$\rm H_2$ cooling and dominates the temperature and density evolution 
of the collapsing gas, and hence its tendency to fragment and 
form new stars. 

Another point to note is that a change in the analysis of the ratio,
$t_{\rm cool}/t_{\rm ff}$, does not necessarily coincide with a change in 
the ratio of the free-fall time to the sound crossing time $\hbox{--}$ 
which is a measure of the number of Jeans masses.
Note that \citet{tao09} were the first to find the fragmentation 
in the Pop~III star forming haloes because they included the heating due 
to $\rm H_2$ formation, whereas \citet{yoh08} did not. These studies 
along with our detailed rigorous analysis support the claims that the 
contribution from both the three-body $\rm H_2$ formation heating and 
compressional heating must be included to capture the fragmentation in 
the Pop~III star forming halos.

\subsection{The formation rate of $\rm H_2$}
Since the onset of the chemothermal instability is strongly 
associated with the formation of hydrogen molecules, it is therefore 
instructive to examine the rate at which $\rm H_2$ actually forms in 
the collapse calculations.

We calculate the $\rm H_2$ formation rate $(dn_{\rm H_2}/dt)$ from 
the data generated in our simulations and plot as a function of 
density in Figure~\ref{fig:gas_props}C. To create these plots, 
we first calculate the number density of hydrogen molecules on 
a mass shell that has a particular density and temperature for a 
specific epoch (i.e., for a given snapshot at time $t_1$). For 
the same snapshot, we then keep on calculating the number density 
of hydrogen molecules for different mass shell (each of which has 
a particular density and temperature). Both the density and 
temperature are radially-logarithmic binned, mass-weighted averages.
We repeat the procedure for the consecutive snapshot of time $t_2$. 
We compare the $\rm H_2$ abundances in these snapshots from the 
simulations over a time interval $\Delta t = t_2 -t_1$, with the 
rates given by
\begin{equation}
\frac{dn_{\rm H_2}}{dt} = \frac{n_{\rm H} (t_2) y_{\rm H_2} (t_2) - n_{\rm H} (t_1) y_{\rm H_2} (t_1)}{t_2 - t_1}.
\end{equation}
Note that an assumption here is made that the number density of the gas 
between the two snapshots does not change significantly and the time 
interval $\Delta t$ is chosen as small as possible ($\approx$ 8 months, 
considerably small compared to both the free-fall and cooling times).

Over the range of densities in which the three-body reactions are important 
(i.e. above $10^7$ cm$^{-3}$), we find that the differences in  
$dn_{\rm H_2}/dt$ between the simulations are strongly density dependent. 
Around a density of $\sim 10^8 \hbox{--} 10^9$ cm$^{-3}$, the difference 
between the formation rates in the simulations is the highest. The
curve is steeper in this regime, allowing us to infer that the rapid 
production of hydrogen molecules can cool the gas promptly and hence 
trigger the cooling time to be shorter than the free-fall time.

Note that at high densities this method is no longer strictly 
applicable as it is highly sensitive to the chosen time interval. However, 
this calculation provides a good estimate for the overall trend of the 
rate of $\rm H_2$ formation inside the haloes in which we are particularly 
interested.

%
%
%
%
%
%
%
%


\begin{figure*}
\centerline{
\includegraphics[width=7.0in]{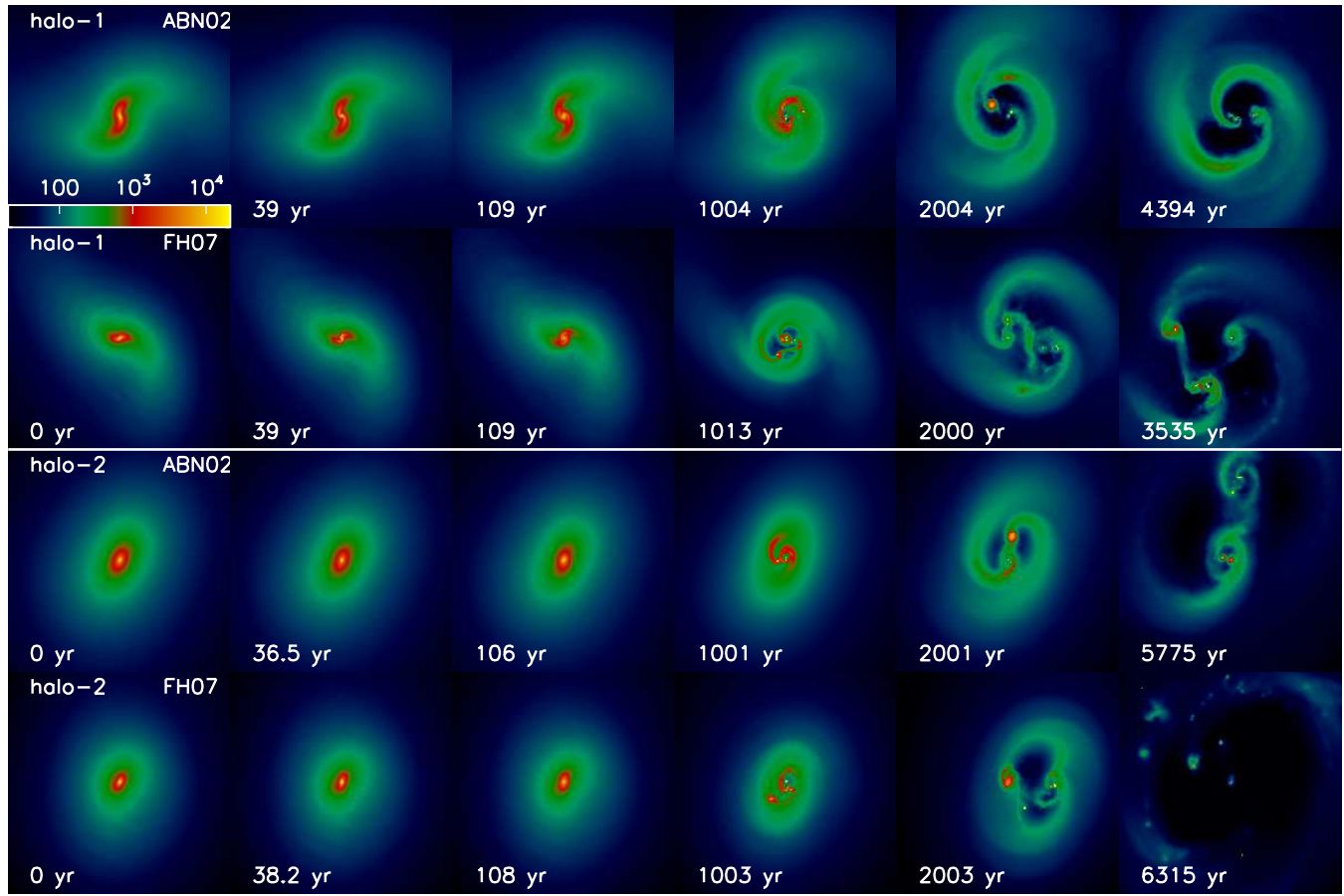}
}
\caption{\label{disk_image} Evolution of surface density in the disk 
at later epochs in a region of 2000 AU centred around first protostar 
in the cosmological minihaloes for different proposed three-body 
$\rm H_2$ formation rates.} 
\end{figure*}

\subsection{Fragmentation and long term evolution}
\label{sec:fragmentation}

Fragmentation can be manifested in various terms, as it 
depends on the thermal and chemical history, interaction with the 
surroundings, gas mass and instability in the clouds. Here we 
focus on how the thermo-dynamical evolution due to the heating 
and cooling rates can provide a hint of future fragmentation in 
our calculations. 

One can measure the instability in the gas by computing 
the number of Jeans mass, or more appropriately the Bonnor-Ebert 
mass ($M_{\rm {BE}}$) inside the central dense volume 
\citep{ebert1955,bonnor1956}. This is same as comparing the 
fragmentation timescale ($t_{\rm frag}$) with the accretion 
timescale ($t_{\rm acc}$). In other words, the ratio, 
$t_{\rm frag}/t_{\rm acc}$, is equivalent to the ratio, 
$M_{\rm BE} / M$, i.e., the inverse of the number of Bonnor-Ebert 
mass masses enclosed in the central volume. We define the fragmentation 
timescale as the rate of change in the number $M_{\rm BE}$ within 
a given radial shell \citep{dgck13}, 
\begin{equation}
t_{\rm frag} \equiv \frac{M_{\rm BE}}{\dot{M}}, 
\end{equation}
and compare it with the accretion timescale \citep{abn02}, defined as 
\begin{equation}
t_{\rm acc} = \frac{M_{\rm enc}(r)}{4 \, \pi \, \rho \, v_{\rm rad} \, r^2},
\end{equation}
where $\dot{M}$, $M_{\rm enc}$, and $v_{\rm rad}$ are the mass 
accretion rate, enclosed mass and radial velocity, respectively. The 
results are shown in Figure~\ref{fig:gas_props}D. The snapshots were 
taken just before the formation of the first protostar. The dashed 
lines represent the case when fragmentation timescale is equal to 
the accretion timescale.

If $t_{\rm frag}/t_{\rm acc} > 1$, the gas enclosed in the shells 
\tr{is} accreted faster than it can fragment. As a result, fewer 
new protostars are formed and the available mass contributes to 
the mass growth of the existing ones. In our case, we find 
$t_{\rm frag}/t_{\rm acc} < 1$, i.e., the gas in the shells can 
fragment faster than it is accreted by the central dense clump, 
favouring low-mass protostars. Therefore, comparing only 
the timescale, we get an indication of the possible 
fragmentation from the thermo-dynamical evolution of the gas 
before the onset of the sink formation.

At later epochs, we follow the simulations for 
$\approx 3500\hbox{--}6500$ yr (depending on the cosmic variance 
of the minihaloes) after the formation of the first sink 
particle. The column density in the inner 2000 AU at the end 
of the simulations are shown in Figure~\ref{disk_image}. In all 
cases we see that the simulations exhibit the disk structure that 
fragments on several scales within this central region. We can 
also see that the density in the disk are slightly less in the 
case with the FH07 rate due to the increased $\rm H_2$ fraction. 
The disk evolves with time with the matter accreted on to 
sink particles, and becomes more and more Jeans unstable. 
  
We conclude that the temperature continues to rise with increasing 
density meaning that there is no preferred scale at which 
fragmentation takes place during the collapse. Any future 
fragmentation is likely to be a result of the drop in the Jeans 
mass that is a strong function of temperature. Hence the intricate 
combination of the heating and cooling rates during the three-body 
reaction plays a significant role in determining the unstable 
clumps in the cloud.


\section{Summary}

We have investigated the diverse cooling and heating mechanism in two 
cosmological minihaloes during the gas collapse to higher density. 
We have compared the physical processes with the extreme three-body 
formation rate coefficients proposed by \cite{abn02} and by \cite{fh07}.
 
We argue that although the quantity $t_{\rm cool}/t_{\rm ff}$ might 
determine the onset of the chemothermal instability, the increase in 
the $\rm H_2$ cooling throughout the fully molecular gas is offset 
by $\rm H_2$ formation heating with compressional heating. Hence the 
thermal instability need not necessarily be the criterion for possible 
fragmentation. Rather, the thermo-dynamical evolution that depends 
largely on the complicated combination of the heating and cooling rates 
develops the unstable clumps by affecting its Jeans mass. This 
leads the disk to fragment on a scale of 1000-2000 AU. However, the 
scale in which the fragmentation takes place depends on the three-body 
$\rm H_2$ formation rates. For example, simulations employing the 
ABN02 rate produce on average fewer and more massive fragments near 
the center, compared to those calculations using the FH07 rate 
\citep{dnck15}. However, the scale of fragmentation is comparable to 
the halo$\hbox{-}$to$\hbox{-}$halo variation. Our conclusion agrees 
well with the study by \citet{ra04} and \citet{yoha06} that pointed that 
although the three-body reaction results in chemothermal instability, 
the growth of fluctuations are too small to lead to any fragmentation.
Despite considerable computational efforts involved, we emphasize 
here that we cannot rule out the plausible concept of the chemothermal 
instability as the sole reason in all Pop~III disk fragmentation
scenarios. This is because of the fact that we have only considered 
two realizations. It is therefore of immediate interest to investigate
the chemothermal instability in a number of minihaloes, like those
in \citet{hirano14}.

In summary, we have explicitly shown various heating and 
cooling rates in a 3D simulation of Pop~III collapse. This unique 
approach to the problem enables us to conclude that the fragmentation 
behavior does not necessarily happen due to the chemothermal 
instability. Instead, fragmentation occurs when $\rm H_2$ cooling, 
the heat released by $\rm H_2$ formation, and $pdV$ compressional 
heating together set the Jeans mass in the gas to values that are 
conducive to breakup and collapse.


%
%
%
%
%

\smallskip
The author is grateful to Prateek Sharma, Agnieszka Janiuk, B. N. 
Dwivedi, Biman Nath, Dominik Schleicher, John Wise and Jarrett Johnson
for throughly checking the manuscript and efficacious comments. The 
author acknowledges the referee for helpful suggestions that have 
helped to ameliorate the clarity of the paper. The author is supported 
by the Indian Space Research Organization grant (No.~ISRO/RES/2/367/10-11) 
to Banibrata Mukhopadhyay and Department of Science and Technology (DST) 
grant (Sr/S2/HEP-048/2012) to Prateek Sharma. The author would also 
like to thank the Department of Physics, Indian Institute of Technology 
(Banaras Hindu University) at Varanasi and the Inter-University Center 
for Astronomy and Astrophysics at Pune for the local hospitality.

\footnotesize{

}
\end{document}